\documentclass{emulateapj}

\shorttitle{Alumina dust in V4332 Sgr}
\shortauthors{Banerjee et al.}

\begin{document}
\title{Spitzer observations of V4332 Sagittarii: Detection of Alumina Dust}

\author{D.\ P.\ K.\ Banerjee\altaffilmark{1},  K.\ A.\
  Misselt\altaffilmark{2}, K.\ Y.\ L.\ Su\altaffilmark{2},
  N.\ M.\ Ashok\altaffilmark{1}, P.\ S.\ Smith\altaffilmark{2}}

\altaffiltext{1}{Physical Research Laboratory, Navrangpura, Ahmedabad
  Gujarat 380009, India. orion,ashok@prl.res.in} 
\altaffiltext{2}{Steward Observatory, University of Arizona, 933 North
  Cherry Avenue, Tucson, AZ 85721. kmisselt,ksu,psmith@as.arizona.edu}

\begin{abstract}
 
We present broad-band 24, 70 and 160~$\micron$ photometry,
5-35~$\micron$ and 55-90~$\micron$ spectra of the eruptive variable
V4332~Sgr from {\it Spitzer} observations. The distinguishing feature
of the 5-35~$\micron$ spectrum is an unusually broad absorption
feature near $10~\micron$ at the position generally associated with
silicate-rich dust. Through radiative transfer modeling, we show that
this broad feature cannot arise from silicates alone but requires the
inclusion of alumina (Al$_2$O$_3$) as a dust condensate. The case for
including Al$_2$O$_3$ is strengthened further by the presence of the AlO
radical, a potentially important molecule in forming Al$_2$O$_3$.  The
present detection indicates that porous alumina manifests itself
through a broadening of the 9.7~$\micron$ silicate feature and
additionally displays, on the shoulder of the silicate feature, a
component at $\sim11.5~\micron$.  We discuss how further observations 
of V4332~Sgr may have the potential of verifying some general 
predictions of the dust condensation process.

\end{abstract}

\keywords{infrared: stars-novae, cataclysmic variables - stars: individual 
(V4332 Sagittarii) }
    
\section{Introduction}

Alumina (Al$_2$O$_3$) is considered to play a significant role in dust
formation around oxygen-rich cool stars. Thermodynamic equilibrium
calculations indicate that it, along with titanium oxides, is one of
the earliest condensates in the mineralogical condensation sequence
\citep{1990fmpn.coll..186T,1999A&A...347..594G}. Observationally, there
is some debate and uncertainty regarding the spectral signatures that
can be ascribed to alumina that permit a firm conclusion to be drawn
for the presence of alumina.  In particular features at 11.3~$\micron$
and 13~$\micron$, seen in the spectra of O-rich AGB and
supergiant stars have often been attributed to alumina
\citep[e.g.,][and references
therein]{2000A&AS..146..437S}. Additionally, various authors have
shown that the inclusion of alumina grains in dust models yields
better fits to the observed profile of the silicate feature at
9.7~$\micron$ (especially when the feature is broad) and also
reproduces better the overall infrared spectral energy distribution
(SED) in selected AGB and OH/IR stars
\citep{2000A&AS..146..437S,2005MNRAS.362..872M,2006ApJ...640..971D}. In this paper, we
present evidence for alumina dust detection from {\it Spitzer Space
Telescope} observations of the nova-like variable V4332 Sgr. The
distinguishing feature of its mid/far IR spectrum is a deep, unusually
broad absorption feature at 10~$\micron$.  We show that this feature
cannot be reproduced by silicate dust alone and that it is necessary
to invoke the presence of amorphous alumina grains to explain it.

V4332~Sgr erupted in 1994 in what was initially considered a nova-like
outburst \citep{1999AJ....118.1034M}.  However, its subsequent
post-outburst evolution to a cool spectral type indicated that this was
not a classical nova eruption. The exact nature of V4332~Sgr is of
considerable interest as it, along with V838 Mon and M31 RV 
(a red-variable which erupted in M31 in 1988), may form a new class of
eruptive objects
\citep[e.g.][]{2002A&A...389L..51M,2003Natur.422..405B}. V4332~Sgr
shows an intriguing emission-line spectrum in the optical and
near-infrared with several rare spectral features. Prominent molecular
bands of TiO, ScO, VO and AlO are also seen in the optical
\citep{2005A&A...439..651T,2006AN....327...44K} implying an oxygen
rich environment. The fundamental band of $^{12}$CO at 4.67$\micron$
has also been detected in the source along with water ice at
3.05$\micron$ \citep{2004ApJ...615L..53B}. The IR excess detected in
the source, along with the molecular and ice features, suggest a cool
dusty environment around the central star whose effective temperature
is estimated to be $\sim$ 3250-3280K
\citep{2003ApJ...598L..31B,2005A&A...439..651T}.

\section{Observations and Data reduction}

V4332~Sgr was imaged with the Multiband Imaging Photometer for {\it
Spitzer} \citep[MIPS;][]{2004ApJS..154...25R} at 24 and 70~$\micron$ on 15 Oct
2005 and 2 Nov 2006 (70~$\micron$ Fine and Default modes,
respectively). Data at 160~$\micron$ were 
obtained on 15 Oct 2005. Spectra were obtained using the Infrared
Spectrograph on {\it Spitzer} \citep[IRS;][]{2004ApJS..154...18H} on 18 April
2005 and 19 April 2006. In 2005, low resolution (R $\sim$ 60-100) data
from $\sim 5-38~\micron$ and high resolution data (R = 600) from $\sim
18-38~\micron$ were obtained. In 2006, high resolution data from $\sim
19-38~\micron$ and low resolution data from $\sim 5-14~\micron$ were
obtained. In addition, MIPS SED mode data covering the wavelength
range from $\sim 55-90~\micron$ were obtained on 27 Sept 2005.  For
the following discussion, data obtained in 2005 and 2006 will be
referred to as epoch 1 and epoch 2, respectively.

\begin{figure}
\plotone{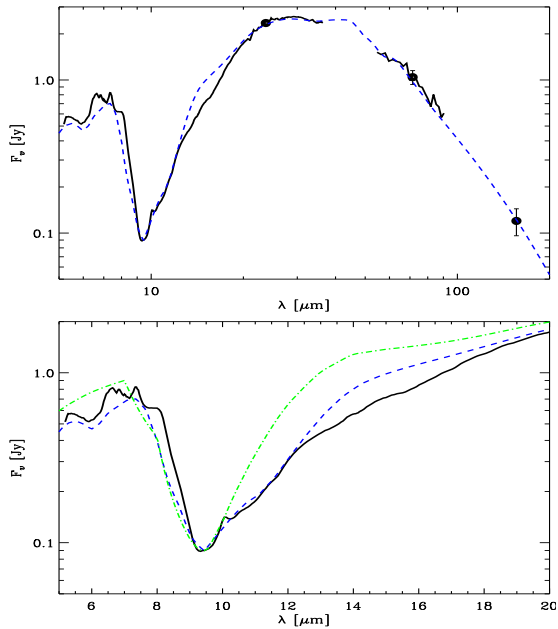}
\caption{Top: Epoch 1 (solid curve and points) along with the best model
  fit (dashed curve); see Table
  \ref{tab:modpar}, column 3. Bottom: Expanded view around the 10~$\micron$
  complex. In addition to the best fit model, the best fit silicate only
  model is over plotted (dash-dotted curve).  It is clearly seen that a pure silicate model
  yields a poor fit to the extended red wing of the
  data. \label{fig:fullSED}}\
\end{figure}

The MIPS data were reduced using the Data Analysis Tool v3.06
\citep{2005PASP..117..503G}. V4332~Sgr was detected as a point source
by MIPS at 24, 70, and 160~$\micron$, and the flux densities were
extracted using both PSF fitting and aperture photometry.  The
measured MIPS flux densities were 2.34$\pm$0.07, 1.07$\pm$0.11, and
0.12$\pm$0.02~Jy at 24, 70, and 160~$\micron$\, respectively.  At 24
and 70~$\micron$\, the flux densities measured in Epochs 1 and 2 were
identical within the errors and we report the weighted mean of those
measurements. The basic instrumental calibration of the MIPS SED mode
is similar to that of the 70~$\micron$ imaging mode
\citep{2007ApJ..InPress} but with a wavelength-dependent illumination
correction. The final spectrum of the source is obtained by extracting
a 5-column (49\farcs25) aperture from the sky subtracted 2-D spectrum
after correcting for slit losses.  The MIPS SED data were scaled by a
factor of 0.88 to match the flux density in the 70~$\micron$ bandpass.
Details of the SED reduction can be found in Lu et al.~(2007, in
prep.). 

The IRS data were processed and extracted using the {\it
Spitzer} Science Center (SSC) Pipeline S15.3 product. Since V4332~Sgr
is a bright point source, the final spectrum of the low resolution
modules were combined using the SSC post-BCD co-added, nod-subtracted
spectra. No background observation was obtained for the high
resolution modules and the background level for the epoch 1 observations
was $\sim$ 0.06 Jy (a factor of 20 fainter than the source) at
$\sim$20~$\micron$ and fairly uniform in the co-added 2-D low
resolution spectra. The SSC post-BCD extraction of the high-resolution
spectrum agrees with the low-resolution spectrum within the
uncertainties. There is no obvious emission/absorption lines seen in
the high-resolution spectrum; therefore, the final spectra of V4332
Sgr in epochs 1 and 2 are computed by averaging both low- and
high-resolution modules. In Figure \ref{fig:fullSED}, we present the
observed IRS and MIPS SED spectra of V4332~Sgr and the combined broad
band MIPS fluxes at 24, 70 and 160~$\micron$.  Since there is little
apparent evolution between the epoch 1 and 2 broad band fluxes, we show
only the epoch 1 data in Figure \ref{fig:fullSED}. Evidence for changes
in the detailed shape of the SED between epochs will be examined
below.

\section{Results}
The spectrum of V4332~Sgr is dominated by a deep, broad feature at
$\sim10~\micron$, normally associated with the presence of amorphous
Mg-Fe silicate grains. However, this observed $10~\micron$ feature is
relatively broad, with an additional feature at $\sim11\micron$ and a
flattened wing beyond $\sim13~\micron$. Additionally, signatures of
ices and organic materials are evident from $\sim5-8~\micron$ (water
ice at 6~$\micron$, ``organics'' at 6.8~$\micron$ and possibly methane
ice at 7.7~$\micron$; see e.g. Bowey \& Hofmeister 2005).  We have
modeled the V4332~Sgr spectrum using the radiation transfer code DUSTY
\citep{1999DUSTYManual}.  The limitations of DUSTY include the
assumption of a spherically symmetric shell of material which may not
be appropriate for V4332~Sgr as the system may have a disk
\citep{2004ApJ...615L..53B}. However since we are interested in
exploring the overall shape of the observed SED rather than providing
a detailed physical model of the complete system, we have restricted
ourselves to the simplest and most generalized assumptions in our
calculations.  As the luminosity of V4332~Sgr is poorly known, we have
fixed the stellar luminosity for V4332~Sgr at 10$^4$~L$_\odot$, the
default input value assumed by DUSTY. This assumption does not affect
the shape of the computed spectrum, only the physical scale of the
system when combined with the dust temperature at the inner radius of
the shell.  We have fit the observed SED with two models viz. model 1
that contains silicate dust only and model 2 with a mixture of
silicate and alumina dust, where the inclusion of alumina is prompted
by the presence of a feature at $\sim11~\micron$ often attributed to
amorphous alumina \citep{2000A&AS..146..437S,2006ApJ...640..971D}.
Prompted by the presence of ice absorption at $\sim6~\micron$, the
grains in both models are coated with an ice mantle (20\% by
volume). The silicate dust optical constants are from
\citet{1984ApJ...285...89D} while the alumina optical data used are
from \citet{1997ApJ...476..199B}.  Corundum was not included in our
model as there is no evidence in our spectra for the feature at
$\sim$13~$\micron$ associated with the presence of this mineral
\citep{2006ApJ...640..971D}.  We tested both the 'porous' and
'compact' alumina samples of \citet{1997ApJ...476..199B}; as there was no
substantive differences between the models, we restricted ourselves to
the 'porous' sample for the subsequent modeling. The ice optical
constants are those of Wiscombe
(ftp://climate1.gsfc.nasa.gov/wiscombe/). The range of parameters
explored is given in Table \ref{tab:modpar}. The output spectra
computed using both dust models are shown in Figure
\ref{fig:fullSED}. It is clearly seen that a pure silicate composition
matches the observed 10~$\micron$ feature poorly. On the other hand,
the inclusion of alumina in model 2 improves the fit
significantly. While there is considerable degeneracy in the fits,
especially between the optical depth and the dust temperature (low
temperature, low optical depth models are somewhat degenerate with
high temperature, high optical depth models, though they consistently
yield formally worse fits), it is notable that no model consisting of
only silicate grains provided a satisfactory fit to the 10~$\micron$
absorption feature. While model 2 provides a good fit overall fit to
the SED and the 10~$\micron$ feature from 9--12.5~$\micron$, it
reproduces neither the flattening beyond $\sim$13~$\micron$, nor the
relatively narrow blue wing. We explored using different silicate
optical constants \citep[e.g.][]{1992A&A...261..567O} as well as
varying the size distribution ($a_{max}$ and $q$) but neither approach
improved the fit in these spectral regions.  It is possible that a
more complex geometry than the simple spherical shell utilized in
DUSTY could improve the fit in these regions. 

\begin{deluxetable}{lll} 
\tablecaption{DUSTY Modeling \label{tab:modpar}}   
\tablehead{ 
\colhead{Parameter} & \colhead{Value}  & \colhead{Best Fit} \\   
}      
\startdata
Stellar Luminosity  & $10^4$L$_{\sun}$ & fixed \\
Stellar Temperature & 3250~K\tablenotemark{1}  & fixed          \\
$R_{out}/R_{in}$    & 1000\tablenotemark{2}  & fixed           \\
Shell $\rho$ Distribution & $r^{-2}$    & fixed \\
Composition         & silicates/alumina       & 65\%/35\%  \\
(fraction by number) & & \\
$\tau_{9.8~\micron}$ &   2 -- 55      & 45 \\
$T_{dust}(R_{in}$)  & 300 -- 1750~K  & 1750~K    \\
Grain Size Distribution & MRN\tablenotemark{3} & fixed \\ 
\enddata
\tablenotetext{1}{Banerjee et al. 2003.}
\tablenotetext{2}{Maldoni et al. 2005.}
\tablenotetext{3}{Mathis et al. 1977; $n(a)\sim a^{-q}$ with
  $a_{min,max}=0.005,0.25~\micron$, $q=3.5$.}
\end{deluxetable}

The plots in Figure \ref{fig:fullSED} permit a few conclusions to be
drawn: (i) the substantial improvement in the fits to the broad
10~$\micron$ feature with the inclusion of alumina indicates that
alumina is being detected in the source and its presence is
manifested by a broadening of the 9.7~$\micron$ silicate feature. A
similar conclusion was reached by \citet{2000A&AS..146..437S} using
data extending to $\leq$13.5~$\micron$.  (ii) A small, yet clearly
discernible, feature is seen at 11~$\micron$. This feature is
attributable to alumina since our model calculations show that
increasing the percentage of alumina in the alumina-silicate mixture
of model 2 enhances the strength of this feature. We note that this
11$\micron$ feature is seen in a significant number of stars studied
by \citet{2000A&AS..146..437S} implying that alumina grains are fairly
prevalent.  

\section{Discussion} 

\subsection{The Case for Alumina Condensation} 
It is perhaps not surprising to see evidence for alumina in the dust
surrounding V4332~Sgr given the presence of the AlO radical in its
optical and NIR spectra
\citep{2003ApJ...598L..31B,2005A&A...439..651T,2006AN....327...44K} and
since AlO can play a critical role in the production of
alumina. Laboratory experiments by \citet{2004A&A...420..547D} show
that aluminum oxide clusters with stoichiometry AlO-(Al$_2$O$_3$)$_n$
are readily formed in laser vaporized metallic Al when quenched in
oxygen and argon.  These clusters are found to be very stable and thus
very good nucleation sites for dust growth. Additionally,
\citet{1995JPChem...99...12225} studied the kinetics of AlO + O$_2$
reactions at temperatures in the range 300-1700K with a view to
studying the fate of aluminum in combustion. The higher end of this
temperature range, it may be noted, is very close to the predicted
condensation temperature \citep[1760K;][]{1990fmpn.coll..186T} of
alumina dust around stars. The \citet{1995JPChem...99...12225}
experiment shows that AlO becomes oxidized to AlO$_2$ by O$_2$. An
additional reaction involving AlO is AlO + O +M = AlO$_2$ + M, where
the ``chaperon'', M, is any atom or molecule present that can remove
some energy from the newly-formed, activated AlO$_2$ (A. Fontijn,
private communication). Newly formed AlO$_2$ can further interact with
AlO to generate alumina: AlO + AlO$_2$ + M = Al$_2$O$_3$ + M.

These results suggest that AlO is likely to play a significant role in
the route to Al$_2$O$_3$ formation. This conclusion has theoretical
support in the work of \citet{1999A&A...347..594G} who show 
that any possible nucleation species that can go on to form dust
around stars should begin with a monomer with exceptionally high bond
energy. The AlO monomer satisfies this criterion and is thus a favored
candidate to lead to the formation of larger Al$_m$O$_n$ clusters that
serve as nucleation sites for the formation of other grains or to
alumina grains themselves by homogeneous nucleation. While the
\citet{1999A&A...347..594G} analysis is based on thermal equilibrium
considerations, an alternative model is the non-equilibrium formation
of chaotic silicates proposed by \citet{1990ApJ...350L..45S} and
\citet{1990Ap&SS.163...79N}. Chaotic silicates form rapidly from a
supersaturated vapor of metal atoms, SiO, AlO and OH in a hydrogen
atmosphere \citep{1990ApJ...350L..45S}. In the initial stages, the
higher reduction of Al with respect to Si will lead to the
preferential formation of Al-O bonds at the expense of Si-O
bonds. This implies that the IR bands of alumina associated with the
Al-O stretching mode should be prominent early in the formation of the
chaotic silicates. However, as the Al atoms become fully oxidized, the
higher abundance of Si will make the 9.7${\rm{\mu}}$m band associated
with Si-O bonds dominate.

Titanium oxides are considered to also be an early dust condensate
along with alumina. Given that the dust that has formed around
V4332~Sgr is of fairly recent origin as implied by the abrupt infrared
brightening that developed in the source between 2MASS observations in
1998 and subsequent observations in 2003 \citep{2003ApJ...598L..31B},
we might expect some signature of these species in the spectra,
though Ti is nearly 30 times less abundant than Al
\citep{2000A&AS..146..437S}. Bulk titanium oxides can have different
forms: TiO, TiO$_2$, Ti$_2$O$_3$ and Ti$_3$O$_5$
\citep{2004A&A...420..547D}. The most common, TiO$_2$ can exist as
brookite and anatase which convert at high temperature into rutile
which is the third and most common form. The rutile spectrum is
expected to show a broad and strong band at 13-17~$\micron$; the
spectrum of anatase shows two strong and broad bands around 17 and
29~$\micron$. The titanium oxide clusters, studied by
\citet{2004A&A...420..547D} as possible nucleation sites, have a
vibrational transition at $\sim$13.5~$\micron$. 
The above discussion pertains mostly to crystalline forms of titanium oxides 
while in V4332 Sgr the amorphous form may  more likely  be present
given that both the silicates and alumina are of  amorphous nature. 
But a reasonable possibility exists that the flattening of the absorption 
band longward of 13.5~$\micron$  is indicating the presence of titanium oxides.

\subsection{Evolution of the Dust Condensates}
\begin{figure} 
  \plotone{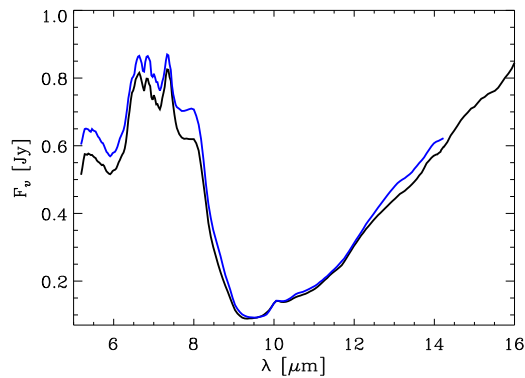}
  \caption{Enlargement of the 10~$\micron$ absorption complex comparing
    epoch 1 (black) and 2 (blue) data. There is evidence for a narrowing of the
    absorption complex between the two epochs.
    \label{fig:epoch_comp}}
\end{figure}

There is potential in the present data to address certain aspects of
the dust condensation process in astrophysical environments. Since the
dust formation  process in V4332 Sgr has begun
recently - certainly less than 10 years ago - and is possibly still
ongoing, there are a few spectral features that could change with
time. As an example, in the ``chaotic silicates'' hypothesis, it is
predicted that a strengthening of the silicate component of
the 9.7~$\micron$ feature should take place relative to that of the
alumina and other early condensate components that blend with this
feature.  There is observational support for such evolution comparing
our data between epochs 1 and 2 (see Fig \ref{fig:epoch_comp}). There
is a hint that the broad red wing of 9.7~$\micron$ feature has
weakened in epoch 2 relative to epoch 1 and that there has been an overall
narrowing of the 10~$\micron$ absorption complex. Although the
evidence is tentative given the small change ($\sim 1 \sigma$) and
only two epochs of data, such a behavior might be expected as Al and
Ti atoms become oxidized and Si-O begins to dominate the composition.
Further, it is also predicted that the ratio of the
10~$\micron$/18~$\micron$ silicate features could be expected to
change monotonically as silicate dust nucleates and anneals in a
circumstellar environment \citep{1990Ap&SS.163...79N}. Freshly
nucleated silicates, as laboratory experiments show, are expected to
have a large 10~$\micron$/18~$\micron$ ratio i.e. the 18~$\micron$
feature is expected to be weak (consistent with what is seen in the
V4332~Sgr spectrum). Thermal processing  should increase the
strength of the 18~$\micron$ feature.  
Although the time scales involved in the above processes are not clear,
it would be worthwhile to monitor the spectrum of V4332 Sgr in the
future to  discriminate between possible scenarios for the evolution of dust
condensates (Dijkstra et al. 2005 and references therein).

The detection of alumina condensate in V4332~Sgr may have implications
for the origin of this object. It has been postulated that, along with
V838 Mon and M31 RV, V4332~Sgr forms a new class of eruptive
variables.  The nature of these objects and the source of their
eruptive behavior has not been established and ideas ranging from an
outburst from a compact object in a red giant envelope, various
thermonuclear events in a single high-mass evolved star, a late
He-shell flash, stellar mergers and even the
swallowing of multiple planets by a massive star have been proposed
\citep[e.g.][and references therein]{2006A&A...451..223T}. Within the
scope of this work, it is not possible to discuss in depth the
complexities of the origin and nature of the V4332~Sgr system. To
date, alumina dust has been almost exclusively detected in AGB and
other cool evolved stars and, as discussed above, is
likely a very early condensate in any oxygen rich environment, so the
detection of alumina in the early condensate of V4332~Sgr indicates
that conditions in the ejecta are similar to those found
around cool evolved stars.  Thus the detection of alumina in V4332~Sgr
may provide a constaint on the nature of the eruption if more detailed
modeling of some of the proposed eruption mechanisms rules out conditions
conducive to the formation of alumina grains.  In addition, the
detection of alumina around V4332~Sgr motivates long-term monitoring
of the ejecta formed around V838 Mon \citep{2006ApJ...644L..57B}. If
indeed these objects are related at a more fundamental level than
simply having roughly similar outburst characteristics, we might
expect that the conditions in the post outburst ejecta of V838 Mon to
be similar to those in V4332~Sgr.  Given that V838 Mon erupted $\sim$8
years after V4332~Sgr and exhibits AlO lines in its spectrum
\citep{2004ApJ...607..460L}, we might detect similar
signatures of alumina formation around V838 Mon in the coming years if
both objects do indeed share a common origin.

\begin{acknowledgements}
Research at the Physical Research Laboratory is funded by the
Department of Space, Government of India. This work is based on
observations made with the {\it Spitzer} Space Telescope, which is operated
by the Jet Propulsion Laboratory, California Institute of Technology
under a contract with NASA. Support for this work was provided by NASA
through Contract Number 1277253 and 1256424 issued by JPL/Caltech.

\end{acknowledgements}


\end{document}